\documentclass[%
 reprint,
 amsmath,amssymb,
 aps,
pra,
]{revtex4-2}

\usepackage{graphicx}% Include figure files
\usepackage{dcolumn}% Align table columns on decimal point
\usepackage{bm}% bold math
\usepackage{xcolor}
\begin{document}

\preprint{APS/123-QED}

\title{Dynamical Signatures of Floquet Topology in Wave Packet Dynamics}% Force line breaks with \\

\author{Xin Shen}%
 \email{shenx@cjlu.edu.cn}
\affiliation{%
College of Sciences, China Jiliang University, Hangzhou 310018, China}%
\author{Bing Lu}
 \affiliation{College of Sciences, China Jiliang University, Hangzhou 310018, China}%Lines break automatically or can be forced 
\author{Yan-Qing Zhu}
\email{zhuyq1992@gmail.com}
\affiliation{Key Laboratory of Atomic and Subatomic Structure and Quantum Control (Ministry of Education), Guangdong Basic Research Center of Excellence for Structure and Fundamental Interactions of Matter, School of Physics, South China Normal University, Guangzhou 510006, China}
\affiliation{Guangdong Provincial Key Laboratory of Quantum Engineering and Quantum Materials,
Guangdong-Hong Kong Joint Laboratory of Quantum Matter,
Frontier Research Institute for Physics, South China Normal University, Guangzhou 510006, China}
\affiliation{Quantum Science Center of Guangdong-Hong Kong-Macao Greater Bay Area, 3 Binlang Road, Shenzhen, China}

\date{\today}% It is always \today, today,

\begin{abstract}
Periodically driven quantum systems, known as Floquet systems, provide a versatile platform for engineering novel topological phases absent in static settings. However, dynamically characterizing these non-equilibrium topological invariants remains a challenge. Here, we develop a Floquet perturbation theory in the extended Hilbert space to analytically describe the center-of-mass (CoM) dynamics of a wave packet. When applied to the driven Su-Schrieffer-Heeger model, our theory reveals that the CoM exhibits multi-frequency Zitterbewegung oscillations, whose spectral composition and phase are directly tied to the system's Floquet band structure. Crucially, we find that band inversions at topological phase transitions imprint distinct signatures in the CoM dynamics, including the emergence of low-frequency modes and phase shifts of the oscillatory trajectory. These dynamical signatures offer a practical protocol for detecting Floquet topological invariants, which we demonstrate for both high-frequency and strongly driven regimes. Our work establishes CoM dynamics as a simple and experimentally accessible probe for exploring topological phase transitions in Floquet systems.
\end{abstract}

\maketitle

%\tableofcontents

\section{Introduction}

The study of periodically driven quantum systems has emerged as a cornerstone in modern condensed matter physics and quantum engineering, offering a unique framework to explore exotic phenomena that are inaccessible in static Hamiltonian systems~\cite{Bukov2015,Bordia2017,Oka2019,Rudner2020,Else2020,Torre2021}. The Floquet theory, originally formulated in the context of classical periodic systems by Floquet in 1883~\cite{Chu1985}, has been extended to quantum mechanics to describe the evolution of systems subjected to time-periodic potentials. This formalism has proven particularly powerful in unraveling the dynamics of periodically kicked systems, such as quantum kicked rotors~\cite{Moore1994,Cao2022,Santhanam2022,Guo2025}, cold atoms in optical lattices \cite{Zheng2014a,Jotzu2014,Tarnowski2019,Wintersperger2020,Eckardt2017, Cooper2019,Weitenberg2021}, and Floquet topological insulators~\cite{ Kitagawa2010, Rudner2013, Nathan2015,Roy2017,Yao2017}.

A critical focus in Floquet physics is the interplay between periodic driving and the emergence of topological properties. Notably, the Floquet framework preserves key aspects of topological classification from static systems~\cite{ Nathan2015,Roy2017,Yao2017}. However, crucial distinctions arise from the system's inherent non-equilibrium nature. In the periodically driven system, those edge states, termed as anomalous edge states, can exist even though the quasienergy bands are topologically trivial~\cite{ Rudner2013}. For the experimental validation of topological systems, recent studies have revealed that the measurements of the topological invariants can be identified through the band inversion surface~\cite{Zhang2020, Zhang2023} or the topological charge at the band-touching points~\cite{Uenal2019}. In the real-space analysis, the Floquet anomalous phase in driven chiral systems has been demonstrated to be amenable to characterization via the local density of state~\cite{Zhang2024}. It is intriguing to note that the dynamics of a quantum particle's center-of-mass (CoM) position can encode essential information about the underlying Floquet topology~\cite{Shen2022,Liang2023}. For instance, the oscillatory behavior of CoM under periodic kicks has been linked to analogues of Zitterbewegung—a quantum phenomenon characterized by rapid oscillations of relativistic particles’ position~\cite{Zawadzki2011}. While these connections have been explored in static systems, their generalization to periodically driven scenarios remains to be explored. 

Central to this work is the development of a Floquet perturbation theory tailored for the CoM dynamics. Traditional Floquet-Magus expansion takes the inverse of frequency as the perturbation parameter~\cite{Magnus1954,Blanes2009, Goldman2014,Eckardt2015,RodriguezVega2021}, which generally leaves out the fact that the amplitude of the CoM dynamics is proportional to the energy gap. To address this limitation, we reformulate the perturbative expansion in the extended Hilbert space in a usual way that the perturbation parameter is proportional to the inverse of the energy gap. This approach not only allows a transparent interpretation of multi-frequency oscillations in CoM dynamics~\cite{ Rusin2013,Junk2020,Reck2020}, but also enables direct comparison with the Floquet-Magnus expansion. By applying this formalism to the driven SSH (Su-Schrieffer-Heeger) model~\cite{Asboth2014,DalLago2015,Fruchart2016}, we demonstrate how the CoM oscillations reveal distinct signatures at Floquet topological phase transitions. Specifically, the emergence of frequency components beyond the driving frequency, akin to static multi-frequency Zitterbewegung~\cite{David2010}, is shown to correlate with topological invariant changes in the Floquet bands. The connection between Floquet dynamics and topological transitions offers a practical implementation. Experimentally, CoM measurements in cold atom setups~\cite{Bloch2008} or photonic lattices~\cite{Ozawa2019} can directly probe these signatures, offering a pathway to identify non-equilibrium topological phases. Theoretically, our analysis bridges the gap between Floquet engineering and dynamical topology by establishing CoM as a versatile probe for phase-sensitive diagnostics. 

This paper is organized as follows. Section \ref{sec2} reviews the Floquet formalism and establishes the theoretical foundation for analyzing CoM dynamics in periodically driven systems. Section \ref{sec3} derives the perturbative expansion framework in the extended Hilbert space and contrasts it with Floquet-Magnus approaches. Section \ref{sec4} applies the formalism to the driven SSH model, detailing numerical simulations of CoM behavior across topological transitions. We further discuss the experimental characterization of topological gap invariants in Sec.~\ref{sec5} and a summary is given in Sec.~\ref{summary}.

\section{Dynamics in periodically driven systems}\label{sec2}
\subsection{Floquet Dynamics }

We first briefly review the foundational aspects of Floquet theory for periodically driven systems. Consider a Hamiltonian $\hat H(t+T)=\hat H(t)$  with period $T$, whose time evolution is governed by the operator 
\begin{equation} \hat U(t, 0) = e^{-i\hat K_F(t)}e^{-i\hat H_Ft},
\label{eq:evolution_operator}
\end{equation}
where  $\hat H_F$ is the Floquet Hamiltonian that dictates long-time dynamics, and $\hat K_F(t)$ is a stroboscopic kick operator with period $T$, encapsulating intracycle micromotion. The decomposition in Eq.~\eqref{eq:evolution_operator} is derived by defining the Floquet Hamiltonian via the stroboscopic condition 
 \begin{equation} 
 \hat U(T, 0) = \hat {\mathcal{T}}\exp\left[-i\int_{0}^{T}\hat H(t)dt\right]\equiv e^{-i\hat H_FT},
 \label{eq:Floquet_Hamiltonian_def} 
 \end{equation}
where $ \hat{ \mathcal{T}}$ is the time-ordering operator, and  the kick operator is defined as 
 \begin{equation}
 e^{-i\hat K_F(t)} \equiv \hat U(t,0)e^{i\hat H_Ft}. 
 \end{equation}
The eigenbasis of $\hat H_F$, denoted by  $|n\rangle$ with eigenvalues $\epsilon_n$, defines the quasienergy spectrum. The corresponding Floquet state is expressed as 
\begin{equation} 
|\psi_n(t)\rangle = e^{-i\epsilon_nt}|u_n(t)\rangle~\text{with}~|u_n(t)\rangle = e^{-i\hat K_F(t)}|n\rangle ,
\label{eq:Floquet_state} 
\end{equation}
 where $|u_n(t)\rangle$  is the Floquet mode with periodicity $T$. The general solution to the Schrödinger equation is then expanded as 
 \begin{equation} \label{psit}
 |\psi(t)\rangle = \sum_n c_n |\psi_n(t)\rangle, \quad \text{with} \quad c_n = \langle n | \psi(0)\rangle, 
 \end{equation} 
 where the completeness of $|n\rangle$ ensures an orthogonal basis.
 
To analyze dynamics in the frequency domain, the Floquet mode is expanded via the Fourier series
\begin{equation} 
|u_n(t)\rangle = \sum_{p=-\infty}^{+\infty} |u_n^p\rangle e^{ip\omega t},
\label{eq:Floquet_mode_Fourier} 
\end{equation} 
where $\omega=2\pi/T$ is the driving frequency. Substituting Eq.~\eqref{eq:Floquet_state} into the Schr\"{o}dinger equation yields the frequency-domain eigenvalue equation 
\begin{equation}
\sum_{p}\left[\hat H_{q-p} + p\omega\delta_{pq}\right]|u_n^p\rangle = \epsilon_n|u_n^q\rangle, 
\label{eq:Floquet_eigen_eq} 
\end{equation} with  denoting the Fourier components of $\hat H(t)$ 
\begin{equation} \hat H_p = \frac{1}{T}\int_0^T \hat H(t)e^{-ip\omega t}dt. 
\end{equation}
A critical consideration is the redundancy in solutions of Eq.~\eqref{eq:Floquet_eigen_eq}. By redefining the quasienergy as $\tilde\epsilon_n = \epsilon_n - p'\omega $, one observes that distinct Floquet states are related via 
\begin{equation} |\psi_n(t)\rangle = e^{-i(\tilde{\epsilon}_n - p'\omega)t}\sum_p |u_n^{p+p'}\rangle e^{ip\omega t}, 
\end{equation} 
indicating that quasienergies are defined modulo $\omega$. The redundant solutions can also be found by examining the frequency-domain eigenequation with 
\begin{equation} \label{EHHam}
\sum_p\left(\hat H_{q-p}+p\omega\delta_{qp} \right)| u^{p+p'}_n\rangle = ( \epsilon_n -p'\omega)| u^{q+p'}_n \rangle. 
\end{equation}
To eliminate the redundancy, we restrict quasienergies to the principal value interval $-\omega/2<\epsilon_n\le \omega/2$. For notational clarity, we rewrite Eq.~\eqref{EHHam} as $\hat {\mathcal H} |nq\rangle = \epsilon_{nq}| nq\rangle   $ in the extended Hilbert space basis, where the eigenstates are column vectors of Fourier components 
\begin{equation} \label{eigVec_EH}
|n0\rangle = \begin{pmatrix} \vdots\\ |u_n^{-1}\rangle \\ |u_n^{0}\rangle \\ |u_n^{1}\rangle \\ \vdots \end{pmatrix}, 
\quad 
|nq\rangle = \begin{pmatrix} \vdots\\ |u_n^{q-1}\rangle \\ |u_n^{q}\rangle \\ |u_n^{q+1}\rangle \\ \vdots \end{pmatrix}, \end{equation}
with eigenvalues  $\epsilon_{nq} = \epsilon_n - q\omega ~(q =0,\pm1 , \pm2,...)$. The Hamiltonian in extended Hilbert space reads 
\begin{equation}
\mathcal {\hat H} = 
\begin{pmatrix}
  &   \vdots   & \vdots & \vdots & \vdots & \vdots &  \\
\dots&\hat H_{1}& \hat H_0 - \omega & \hat H_{-1 }&\hat H_{-2} & \hat H_{-3} & \dots\\
\dots&\hat H_2& \hat H_1&\hat H_0  &\hat  H_{-1} & \hat H_{-2} & \dots\\
\dots&\hat H_{3}& \hat H_{2}&\hat H_{1}  & \hat H_0 +\omega &\hat  H_{-1} & \dots  \\
  &   \vdots   & \vdots & \vdots & \vdots & \vdots &  \\
\end{pmatrix}
\end{equation}
The orthonormality condition $\langle mp|nq\rangle=\sum_i \langle u_m^{i+p}| u_n^{i+q}\rangle =\delta_{mn}\delta_{pq}$ ensures proper normalization of Floquet modes.

 The Floquet Hamiltonian  $\hat H_F$ and the stroboscopic kick operator $\hat K_F(t)$ can be systematically derived through the Floquet-Magus expansion \cite{Magnus1954}, which yields leading-order terms\begin{eqnarray}\label{HandK}
\begin{split}
\hat H_F(t_i) &\approx \hat H_0 + \sum_{p\neq 0} \frac{ \hat H_p\hat H_{-p}+e^{ip\omega t_i}[\hat H_0,\hat H_p] } {p\omega},\\
\hat K_F(t) & \approx \sum_{p\neq 0 } \frac{i\hat H_p}{p\omega}\left( e ^{ip\omega t_i} -  e ^{ip\omega t} \right),
\end{split}
\end{eqnarray}
where the intial time $t_i$ is fixed at $t_i = 0$ in our analysis.  
 In Sec.~\ref{sec3} B, we focus on stroboscopic dynamics under the condition where the kick operator $\hat K_F$ vanishes identically, rendering the Floquet Hamiltonian $\hat H_F$ as the sole generator of time evolution. By combining the Floquet-Magnus framework and conventional perturbation theory based on the extended Hilbert space formalism, we demonstrate that the standard approach appropriately reproduces both the energy gap dependence of the oscillation frequencies and the amplitude scaling of the CoM dynamics in the quasienergy basis. This establishes the validity of perturbative approximations in characterizing the multi-mode Zitterbewegung phenomenology observed in our system.

\subsection{Dynamics of the Center-of-Mass}
To analyze the system's macroscopic dynamics, we examine the CoM position $\langle \hat x(t)\rangle = \langle \psi(t)| \hat x| \psi(t)\rangle$, where $\hat x$ is the coordinate operator. By substituting the general solution Eq.~\eqref{psit} into the definition of $\langle \hat x(t)\rangle$, we have
 \begin{equation}\label{xt} 
 \langle \hat x(t)\rangle = \sum_{m,n,p,q} c_m^*c_n \langle u_m^q | \hat{x} | u_n^p \rangle e^{i(\epsilon_{mq} - \epsilon_{np})t}. 
 \end{equation} 
This analysis reveals the emergence of multimode Zitterbewegung \cite{Rusin2013}, wherein the oscillation frequencies are determined by quasienergy band gaps within the extended Hilbert space framework. Specifically, in terms of the eigenvectors and eigenenergies of the extended Hamiltonian $\mathcal H$, the CoM position expectation value can also be expressed as
 \begin{equation}\label{xt2}
 \langle\hat{x}(t)\rangle = \sum_{m,n} \sum_p c_m^*c_n \langle m0 | \hat{x} \otimes \hat{\mathbb{I}} | np \rangle e^{i(\epsilon_{m0} - \epsilon_{np})t},
 \end{equation}
where   $\hat{\mathbb I}$ is the identity operator on the frequency subspace. This expression clarifies that the oscillatory dynamics originate from coherent superposition of quasienergy states, with the frequency spectrum directly tied to the quasienergy level spacing. The observed multimodal oscillations therefore constitute a dynamical manifestation of band interference \cite{David2010}, paralleling the multi-frequency Zitterbewegung phenomena in static band structures while extending their applicability to periodically driven systems. 

\section{Perturbation theory for the Floquet dynamics }\label{sec3}
In this section, we construct a perturbative expansion for the Floquet state,  
\begin{equation} |\psi(t)\rangle = \sum_{n,p} c_n e^{-i\left\{ \left[\sum_k\epsilon_n^{(k)}\right] - p\omega  \right\}t}\left[ \sum_k |u_n^p\rangle^{(k)} \right], 
\label{Floquet_expansion} 
\end{equation} 
where the superscript $(k)$ denotes the $k$-th order correction to the quasi-energy  $\epsilon_n$ and Floquet state $|u_n^p\rangle$ .
\subsection{Perturbation in the extended Hilbert space}
For monochromatic driving Hamiltonians of the form 
\begin{equation} \hat H(t) = \hat H_0 + \hat V e^{i\omega t} + \hat V^\dagger e^{-i\omega t}, 
\label{monochromatic_Hamiltonian} 
\end{equation} 
the system's dynamics is governed by the extended Hamiltonian~\cite{Shirley1965,Sambe1973}, 
\begin{equation}\hat  {\mathcal{H} }= \hat{  \mathcal{H} }_0 + \hat { \mathcal{V} }, 
\label{extended_Hamiltonian} 
\end{equation} where 
\begin{eqnarray}
\begin{split}
\hat{\mathcal H}_0 =& \begin{pmatrix}
\ddots& & & & \\
 &\hat H_0-\omega &  & & \\
 & &\hat H_0 &  & \\
 & & &\hat H_0+\omega & \\
 & & & & \ddots\\
\end{pmatrix}
\\
\hat {\mathcal V} = &\begin{pmatrix}
\ddots& & & & \\
\hat V &0 & \hat  V^\dagger& & \\
 &\hat V &0&\hat V^\dagger & \\
 & &\hat  V&0 &\hat  V^\dagger\\
 & & & & \ddots\\
\end{pmatrix}.
\end{split}
\end{eqnarray}This formulation explicitly captures the frequency-space coupling between Floquet sidebands induced by the time-periodic term.
Under the assumption $|\hat V|\ll \Delta$, where $\Delta$  represents typical band gaps of $H_0$, the perturbation series can be truncated at the first-order correction for single-photon processes. Defining $\tilde \epsilon_n^{(0)}=\epsilon_n^{(0)}+ \mathcal Z(n)\omega$, where $\mathcal Z(n)$ is a integer such that  $\tilde \epsilon_n^{(0)}\in(-\omega/2,\omega/2]$, we obtain the first-order correction to the Floquet eigenstates
\begin{equation} \label{higher_order_terms}
\begin{split}
  |u_n^0\rangle^{(1)} &= |\nu_n\rangle,  \\
  |u_n^{-1}\rangle^{(1)} &= \sum_{m \neq n} \frac{\langle \nu_m|\hat V^\dagger|\nu_n\rangle}{\epsilon_n^0 - (\epsilon_m^0 - \omega)}|\nu_m\rangle,  \\
  |u_n^{+1}\rangle^{(1)} &= \sum_{m \neq n} \frac{\langle \nu_m|\hat V|\nu_n\rangle}{\epsilon_n^0 - (\epsilon_m^0 + \omega)}|\nu_m\rangle, \\
   |u_n^p\rangle^{(1)} &= 0 \quad \text{for } |p| \geq 2,
   \end{split} 
 \end{equation}
where the state $|\nu_n\rangle$ denotes an eigenstate of $\hat H_0$ with eigenvalue $\epsilon_n^0$, i.e., $\hat H_0|\nu_n\rangle = \epsilon_n^0 |\nu_n\rangle$.  From the definition of  $|n\rangle $ and $|u_n^p\rangle$ given in Eqs.~\eqref{eq:Floquet_state} and \eqref{eq:Floquet_mode_Fourier}, we have 
\begin{equation} \label{nkx}
|n\rangle = \sum_p| u_n^p \rangle .
\end{equation}
The physical-space wavefunction then becomes 
\begin{equation}\label{n_EH}
 |n\rangle_{\text{EH}}^{(1)} = |\nu_n\rangle + \sum_{m \neq n}\left[\frac{\langle \nu_m|\hat V|\nu_n\rangle}{\epsilon_n^0 - (\epsilon_m^0 + \omega)} + \frac{\langle \nu_m|\hat V^\dagger|\nu_n\rangle}{\epsilon_n^0 - (\epsilon_m^0 - \omega)}\right]|\nu_m\rangle. 
\end{equation}

\subsection{Comparison with the Floquet-Magnus expansion}
The perturbative expansion presented above is constructed in the order of $\hat V$, contrasting with the Floquet-Magnus expansion formulated in the order $1/\omega$. In the context of stroboscopic dynamics for the CoM, we set  the final time $t=NT$ , where $N=0,1,2,3...$. The time-evolution operator simplifies as
 \begin{equation}
 \hat U(t, 0) = [\hat U(T, 0)]^N = e^{-i \hat H_F N T}.
 \end{equation} 
 where $\hat H_F$ is the Floquet Hamiltonian. To compare with the results obtained via the extended Hilbert space approach [Eq.~(\ref{n_EH})], we derive the first-order corrections to the eigenvalues and eigenstates of $\hat H_F$. By designating  $\hat H_0$ as the unperturbed Hamiltonian and  $ \sum_{p\neq 0} [\hat H_0,\hat H_p]/p\omega $ as the perturbation term, we confirm that the first-order energy shift vanishes, consistent with expectations. For the eigenstate approximation, we obtain
 \begin{equation}\label{n_Floquet} |n\rangle^{(1)}_{\text{FM}} = |\nu_n\rangle + \frac{1}{\omega} \sum_{m \neq n} |\nu_m\rangle \left[ \langle \nu_m | \hat V^\dagger - \hat V | \nu_n \rangle \right]. \end{equation}
In the high-frequency limit $\omega\gg \Delta\epsilon$, Eq.~(\ref{n_Floquet}) converges to a form nearly identical to that of the extended Hilbert space approach [Eq.~(\ref{n_EH})], expressed as
 \begin{equation}\label{n_EH2} |n\rangle^{(1)}_{\text{EH}} = |\nu_n\rangle + \frac{1}{\omega} \sum_{m} |\nu_m\rangle \left[ \langle \nu_m | \hat V^\dagger - \hat V | \nu_n \rangle \right]. \end{equation}
 The sole distinction lies in the inclusion of all intermediate eigenstates in the extended Hilbert space formulation. This discrepancy arises due to differing normalization conventions: the Floquet-Magnus expansion enforces the standard perturbation-theoretic normalization $\langle \nu_n|n\rangle=1$ , whereas the extended Hilbert space approach does not inherently impose this condition. Enforcing the same normalization $\langle \nu_n|n\rangle_{\text{EH}}^{(1)}=1$  on the extended Hilbert space result [Eq.~(\ref{n_EH2})] yields Eq.~(\ref{n_Floquet}), demonstrating that the Floquet-Magnus expansion constitutes a further high-frequency approximation of the standard perturbation theory when the driving frequency exceeds static energy gaps by a significant margin.

\subsection{Heisenberg's equation of motion of $\hat x$}

In the framework of perturbation theory, the non-unitary evolution of the CoM arises due to the non-normalized character and non-orthogonality of perturbed eigenstates in the extended Hilbert space. To rectify this, we impose a consistency condition requiring the Heisenberg equation of motion for the coordinate operator $\hat x$ to hold identically with the dynamical prescription of Eq.~\eqref{xt}. This necessitates that the expectation value $\langle\hat  x\rangle$  obeys
 \begin{equation}
  i\frac{d}{dt}\langle \hat x(t)\rangle = \langle \psi(t) | [\hat{x}, \hat{H}(t)] |\psi(t)\rangle ,
 \end{equation} where the state $|\psi(t)\rangle$ is defined in Eq.~\eqref{psit} and $\hat H(t)$ is the time-periodic Hamiltonian. For an external drive  independent of momentum, substituting Eqs.~\eqref{psit} and \eqref{xt} into the above equation yields the identity 
 \begin{equation}\label{comm_identity}
 \sum_l\langle u_m^{l+q}| \hat x| u_n^{l+p}\rangle = \sum_l\frac{\langle u_m^{l+q}| [\hat x, \hat H(t)]| u_n^{l+p}\rangle }{\epsilon_{np}-\epsilon_{mq}}
 \end{equation} 
 valid for  $mp\neq nq$. Eq.~\eqref{comm_identity} can, in fact, be derived more simply in the extended Hilbert space, using a correspondence analogous to the static case~\cite{David2010}. Using the commutation relation 
 \begin{equation}
[\hat x\otimes \hat{\mathbb I}, \hat {\mathcal H}] = i\frac{\partial}{\partial \hat p_x}\hat H_0\otimes \hat {\mathbb{I} },
\end{equation}
 where $\hat {\mathcal H}$ is the extended Hamiltonian, we have 
\begin{equation}
\langle mq| [ \hat x\otimes \hat{\mathbb I}, \hat {\mathcal H}] | np\rangle = \langle mq| i\frac{\partial}{\partial \hat p_x}\hat H_0\otimes \hat {\mathbb{I} }  | np\rangle. 
 \end{equation}
Consequently, 
\begin{equation}
\langle mq|  \hat x\otimes \hat{\mathbb I} | np\rangle(\epsilon_{np}-\epsilon_{mq}) = \langle mq| i\frac{\partial}{\partial \hat p_x}\hat H_0\otimes \hat {\mathbb{I} }  | np\rangle ,
 \end{equation} 
 where $\hat{\mathcal H}|np\rangle=\epsilon_{np}|np\rangle $ is used. 
 Expanding the matrix product  with the eigenvectors given in Eq.~\eqref{eigVec_EH} yields  \begin{equation}\label{comm_identity_EH}
  \sum_l\langle u_m^{l+q}| \hat x| u_n^{l+p}\rangle = \sum_l\frac{\langle u_m^{l+q}| i\frac{\partial}{\partial \hat p_x}\hat H_0  | u_n^{l+p}\rangle }{\epsilon_{np}-\epsilon_{mq}}.
 \end{equation}
Equation \eqref{comm_identity_EH} can be identified with Eq.~\eqref{comm_identity} by noting that $[\hat x, \hat H(t)]=i\frac{\partial}{\partial {\hat p_x}}\hat H_0$, which follows from the assumption that the external drive is momentum-independent. A similar relation can, however, be derived even when the drive depends on momentum.

Substituting the identity into Eq.~\eqref{xt}, the CoM expectation becomes
\begin{equation} \label{xt2}
\langle \hat x(t)\rangle = \sum_{mq, np } c_m^*c_n \frac{ \langle u_m^q|  [\hat  x,\hat H(t)]| u_n^p\rangle}{ \epsilon_{np}- \epsilon_{mq} } e^{i(\epsilon_{mq}-\epsilon_{np})t}
\end{equation}
 ensuring strict compliance with the Heisenberg's equation through the relation
 \begin{equation}  \label{xt_v}
 \frac{d}{dt}\langle \hat x(t)\rangle = \langle \psi(t)| \hat{v} |\psi(t)\rangle \quad \text{where} \quad \hat{v} = -i[\hat{x}, \hat{H}] . \end{equation}
Under the perturbative expansion, the norm of the evolved state
 \begin{equation} \langle \psi(t)|\psi(t)\rangle = \sum_{m,n,p,q} c_m^*c_n e^{i(\epsilon_m-\epsilon_n+q\omega)t} \langle u_m^p | u_n^{p+q}\rangle
 \end{equation}
 is maintained to first order in $\hat V$ due to the orthogonality condition
 \begin{equation}
 \label{orthogonality}
  \sum_p{ ^{(1)}\langle u_m^p | u_n^{p+q}\rangle^{(1)} } = 
  \begin{cases} 
  \delta_{mn} & \text{for}~q=0 , \\ 0 & \text{otherwise},
  \end{cases} 
  \end{equation}
 derived from the perturbed eigenstate structure in Eq.~\eqref{higher_order_terms}. This ensures $\langle \psi(t)|\psi(t)\rangle=1+\mathcal O(\hat V^2)$ and hence 
 \begin{equation}
 \langle \hat x(t)\rangle =\frac{ \langle \psi(t)| \hat x|\psi(t)\rangle }{ \langle \psi(t) |\psi(t)\rangle}= \langle \psi(t)|\hat x|\psi(t)\rangle
 \end{equation}
 to the first order of $\hat V$.

\section{model}\label{sec4}
\begin{figure}[t]
\centering
\includegraphics[width = 8.5cm]{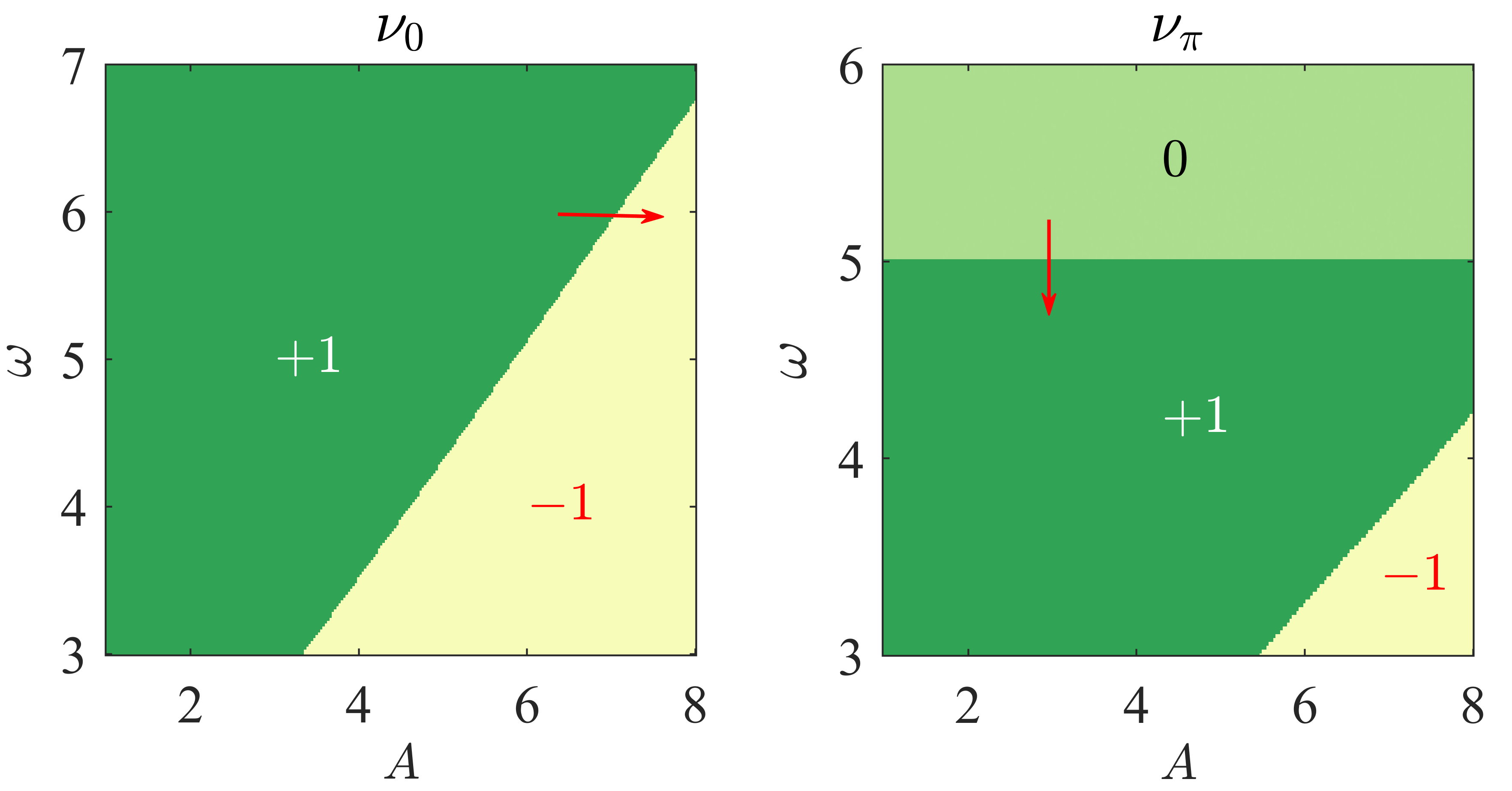}% Here is how to import EPS art
\caption{\label{fig_phase_diagram}  Driving induced Floquet topological phase transition. The chiral invariants are calculated through Eq.~\eqref{wn} for $J_1=1$ and $J_2=1.5$. $J_1$ is taken as the unit for both horizontal and vertical axes.  }
\end{figure}

In this section,  we analyze a driven Su-Schrieffer-Heeger (SSH) model within the framework of perturbation theory to study driving-induced Floquet topological transitions. The SSH model, a prototypical one-dimensional tight-binding lattice with dimerized hoppings, has long served as a playground for understanding static topological phase transitions. When subjected to periodic driving, the SSH model evolves into a Floquet system where the interplay between driving frequency and parameter asymmetry can induce novel topological transitions~\cite{DalLago2015,Fruchart2016}. The time-dependent lattice Hamiltonian is formulated as 
\begin{equation}
\hat H(t) = \sum_i J_1(t)\hat c^\dagger_{iA}\hat c_{iB}+ J_2\hat c^\dagger_{iA}\hat c_{i+1,B}+\text{H.c.},
\end{equation}
where $\hat c_{iA/B}$ is the annihlation operator on the $A/B$ sublattice and H.c. stands for the Hermitian conjugate terms. The intra-cell and inter-cell hopping parameters are $J_1(t)=J_1+A\cos(\omega t)$ and $J_2$, respectively. In the Bloch basis $(c^\dagger_{k_xA}, c^\dagger_{k_xB})$,  with the lattice constant being set to unity, the instantaneous Hamiltonian assumes the form
\begin{eqnarray}\label{hk}
h(k_x,t) = \left[ J_1(t)+J_2\cos(k_x) \right]\sigma_x+J_2 \sin(k_x)\sigma_y,
\end{eqnarray}
with $k_x$ being the quasimomentum and $\sigma_{x,y,z}$ the Pauli matrices. The time-dependent Hamiltonian has the chiral symmetry in the sense that $\sigma_z h(k_x,t)\sigma_z = -h(k_x,-t)$, and together with time-reversal and particle-hole symmetries, it belongs to the BDI symmetry class. 
The topological characterization is achieved through the Floquet invariants  $\nu_\epsilon\in\mathbb Z\times \mathbb Z$ defined as~\cite{Asboth2013,Asboth2014,Fruchart2016,Roy2017,Yao2017}
\begin{equation} 
\nu_\epsilon = \frac{i}{2\pi} \int_{\text{BZ}} dk_x  (V_\epsilon^+)^{-1} \partial_{k_x} V_\epsilon^+, 
\end{equation} 
where $\epsilon=0,\pi$ corresponds to quasienergy gaps and $V_\epsilon^+$ is the evolution operator's periodicized matrix element incorporating half-period dynamics~\cite{Fruchart2016}, i.e., the non-vanishing matrix element of  
\begin{equation} \label{wn}
U^{k_x}_\epsilon\left(\frac{T}{2},0\right) =U^{k_x}\left(\frac{T}{2},0\right)e^{ih_\epsilon^{\text{eff}}\frac{T}{2}}. 
\end{equation}
 Here, the effective Hamiltonian 
 \begin{equation}
h_\epsilon^{\text{eff}} = \frac{i}{T} \log_{-\epsilon}U^{k_x}(T,0),
\end{equation}
with $U^{k_x}(t,0) = \hat{\mathcal T}\exp\left[-i\int_{0}^{t} h(k_x,t)dt\right]$,  accounts for branch-cut considerations in the Floquet operator. The invariant $\nu_0$ reduces to the static topological invariant in the high-frequency limit, while $\nu_\pi$ uniquely captures anomalous edge states absent in static systems~\cite{Yao2017}. 
The phase diagram in Fig.~\ref{fig_phase_diagram} demonstrates driving-induced transitions governed by these invariants. By tuning the driving amplitude $A$  and frequency $\omega$, the system undergoes phase transitions marked by quasienergy band closures at either the $0$ or $\pi$ gap. The study further explores the dynamical response of the driven system, linking Floquet band inversions to the CoM dynamics, as detailed in subsequent sections.

To systematically analyze the driven dynamics of the Floquet SSH model, we initiate the system with a spin-polarized Gaussian wave packet, a configuration experimentally realized in ultracold atomic systems~\cite{Hasan2022}. In free space, the initial state is defined as \begin{equation}\label{eq:initial_state} 
|\psi(t=0)\rangle = \frac{1}{\sqrt{d\sqrt{\pi}}} e^{-\frac{x^2}{2d^2}+ik_x^0x} \cdot \begin{pmatrix} 1 \\ 0 \end{pmatrix}, \end{equation} 
where the spatial Gaussian profile with width $d$  ensures a narrow momentum distribution centered at $k_x^0$ in reciprocal space. This state corresponds to a localized wave packet in real space with a well-defined initial momentum $k_x^0$, serving as an ideal probe for studying Floquet-Bloch dynamics.

For analytical consideration, we focus on the critical point $k_x^0$  where band inversion occurs under time-periodic driving~\cite{Roy2017,Yao2017}. Expanding around this symmetry point $(q_x = k_x-k_x^0)$, the system's dynamics are governed by an effective Hamiltonian
\begin{equation}\label{eq:heff} 
h_{\text{eff}}(t) = \left[J_1(t) + J_2\right]\sigma_x + J_2 q_x \sigma_y
\end{equation} 
 and hence the associated velocity operator is derived as $\hat v = J_2\sigma_y$. By applying Eqs.~\eqref{Floquet_expansion} and \eqref{higher_order_terms}, to the first order, the time-evolved state assumes the form 
\begin{equation}
|\psi(t)\rangle = \sum_{\pm} c_{\pm}e^{-i\epsilon_\pm^0 t} \left[ 1 \mp  i\frac{A}{\omega}\sin(\omega t) \right]|x,\pm\rangle, 
\end{equation}
with eigenbasis $| x,\pm\rangle =1/\sqrt{2}(1,\pm 1)^T$  and unperturbed eigenenergies $\epsilon_{\pm}^0 = \pm(J_1+J_2)$. The coefficients $c_\pm = \langle x,\pm | \Phi \rangle$, where $|\Phi\rangle=(1, 0)^T$ denotes the initial spinor part, quantify the initial state's overlap with Floquet eigenmodes.
The time evolution of the CoM position is obtained by direct expectation value calculation with Eq.~\eqref{xt_v}
\begin{equation} 
\frac{d}{dt}\langle \hat x \rangle = -J_2 \sin(2\epsilon_+^0 t) - \frac{2AJ_2}{\omega}\cos(2\epsilon_+^0 t)\sin(\omega t). 
\end{equation}
 Accordingly,  the first-order corrected CoM is 
\begin{eqnarray} \label{xt3}
\begin{split}
\langle \hat x(t) \rangle_{\text{1st}} = &\frac{J_2}{2\epsilon_+^0}\left[ \cos(2\epsilon_+^0 t)-1\right]\\
-&\frac{A}{\omega}\times \frac{J_2}{2\epsilon_+^0-\omega}\left\{\cos\left[(2\epsilon_+^0-\omega)t\right]-1\right\}\\
+&\frac{A}{\omega}\times \frac{J_2}{2\epsilon_+^0+\omega}\left\{\cos\left[(2\epsilon_+^0+\omega)t\right]-1\right\}.
\end{split}
\end{eqnarray}
The emergent CoM dynamics in Eq.~\eqref{xt3} reveals the interplay between intrinsic band physics and Floquet driving effects. Specifically, the first term originates from the natural Zitterbewegung oscillation dictated by the band gap $\Delta\epsilon = 2\epsilon_+^0$, which manifests the $1/\Delta\epsilon$ amplitude dependence and $\Delta\epsilon$-proportional frequency characteristic~\cite{Zawadzki2011}. The second and third terms, arising from driving-induced sidebands, show a similar structure to the first term, while their amplitudes scale as $A/\omega$, indicating the perturbative origin. 

\begin{figure}[t]
\includegraphics[width = 8.5cm]{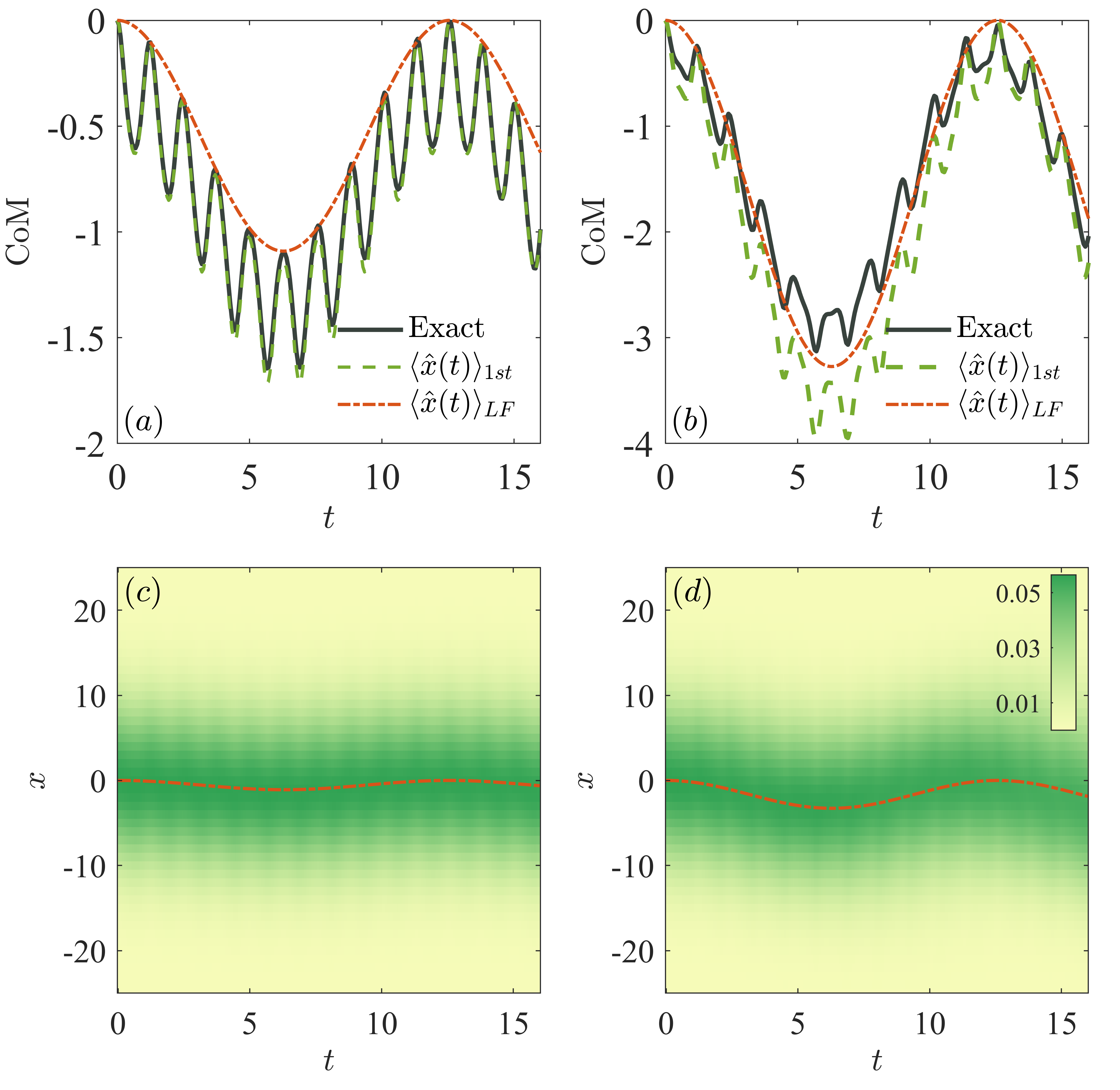}
\caption{\label{fig_xt_diagram} The evolution of the wave packet's center of mass governed by the Hamiltonian Eq.~\eqref{hk}.  The solid lines represent exact numerical solutions obtained by directly calculating the time-dependent Schr\"{o}dinger's equation, while the dashed lines correspond to results from the first-order perturbative approximation of the wavefunction. The dot-dashed curves depict the low-frequency component extracted from the first-order approximation. All calculations employ the parameters $ J_1 = 1 $,  $ J_2 = 1.5 $, $ \omega = 5.5 $, and the width of wave packet $d = 10 $. Panels (a) and (b) respectively use $A = 1$ and $A = 3 $ as the driving amplitude. The time is in units of $1/J_1$ and the length is in units of the lattice space.   (c) and (d) are the spatial-temporal density profiles of the wave packets corresponding to (b) and (d), respectively.}
\end{figure}
Generally, Zitterbewegung dynamics can be expressed as a superposition of interband interference terms. In our case, the first term on the right-hand side of Eq.~\eqref{xt3} stems from the interference between the $\epsilon_+$ and $\epsilon_-$ bands. The second term corresponds to the interference involving the sidebands  $(\epsilon_+, \epsilon_-+\omega) $ or  $(\epsilon_+-\omega, \epsilon_-) $. Similarly, the last term results from the bands  $(\epsilon_+, \epsilon_--\omega) $ or $(\epsilon_++\omega, \epsilon_-) $. When the energy gap for any of these interference terms becomes small, the associated oscillation amplitude grows large. Consequently, the CoM dynamics will be dominated by the corresponding low-frequency component. This is illustrated in Fig.~\ref{fig_xt_diagram} for the case where $|2\epsilon_+^0-\omega|\ll |2\epsilon_+^0 |,|2\epsilon_+^0+\omega| $
and the low-frequency component is 
\begin{eqnarray}\label{xt_LF}
\langle \hat x(t)\rangle_{\text{LF}} = -\frac{A}{\omega} \frac{J_2}{2\epsilon_+^0-\omega}\left\{\cos\left[(2\epsilon_+^0-\omega)t\right]-1\right\}.
\end{eqnarray}

\begin{figure}[t]
\includegraphics[width = 8.5cm]{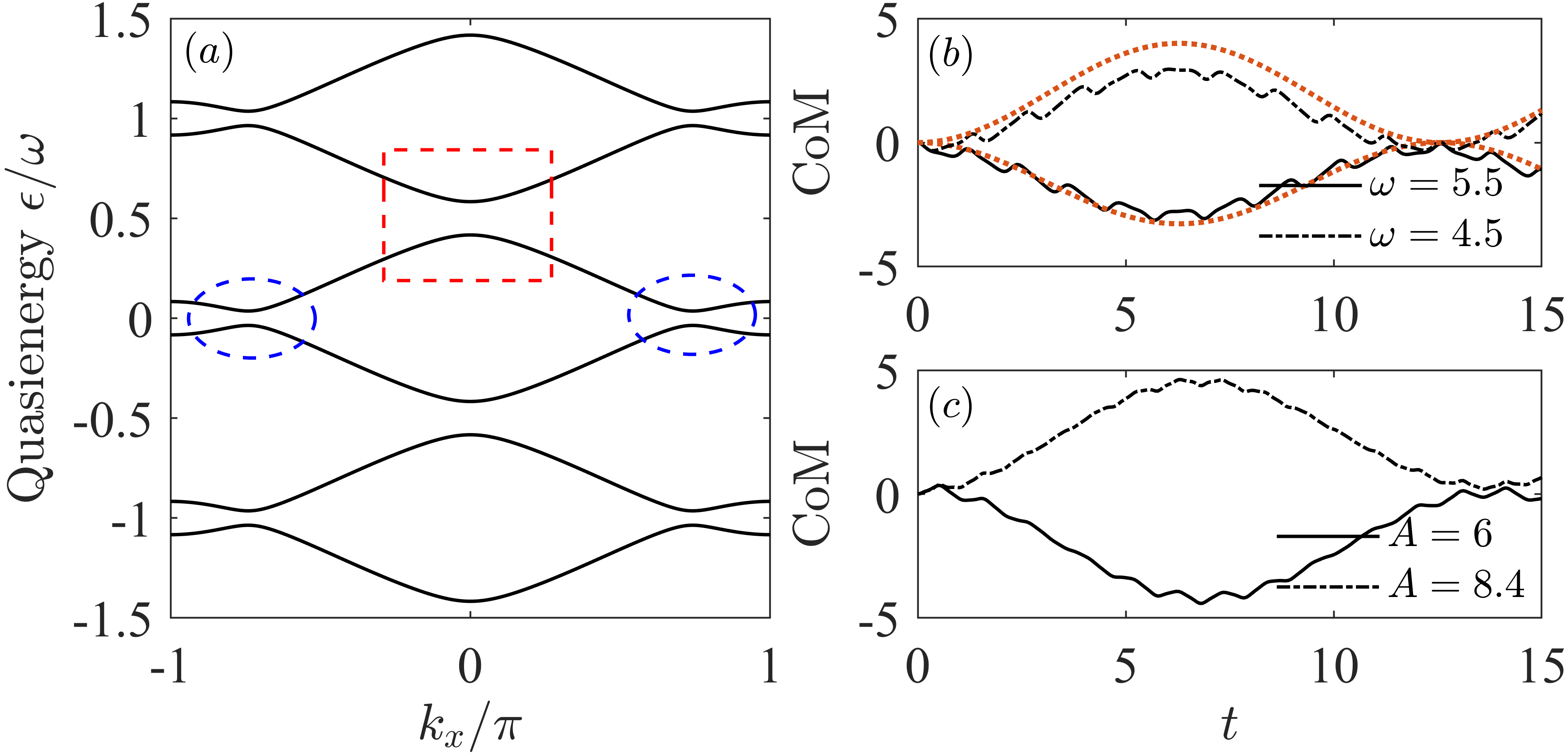}
\caption{\label{fig_phase_transition}  (a) Typical quasienergy spectrum obtained via diagonalization of the extended Hamiltonian. The dashed square and circles highlight the band inversion points within the $\pi$-gap and $0$-gap regions, respectively. In panels (b) and (c), solid lines and dot-dashed curves represent the exact numerical solutions. In (b), the frequency-induced phase transition is marked by the arrow in the right panel of Fig.~\ref{fig_phase_diagram}. The dotted lines correspond to the low-frequency components calculated by Eq.~\eqref{xt_LF}. In (c), the amplitude-induced phase transition is indicated by the arrow in the left panel of Fig.~\ref{fig_phase_diagram}, which lies outside the perturbative regime defined by $A/\omega\ll 1$.  The parameters are $A=3, k_x^0=0$ in (b) and $\omega = 6,k_x^0=\arccos(J_1/J_2) $ in (c). }
\end{figure}
It is noteworthy that the CoM dynamics can identify the band inversion and therefore characterize the Floquet topological phase transition. In analogy to the static case, the transition occurs when the quasienergy gap closes and reopens. This process leads to a local inversion of the energy sequence of eigenstates at the transition point, resulting in distinct dynamical behavior. For instance, consider the oscillatory mode induced by the pair of bands $\epsilon_\pm$, corresponding to the first term in Eq.~\eqref{xt3}. The band inversion reverses the sign of $\Delta \epsilon = \epsilon_+ - \epsilon_- = 2\epsilon_+$, while the magnitude of the gap may differ before and after the phase transition. Consequently, the CoM dynamics acquire an overall negative sign following the transition, indicating a reversal in the direction of CoM vibration. The same principle applies to the second and third terms in Eq.~\eqref{xt3}. Figure~\ref{fig_phase_transition} presents the  CoM position as a function of time before and after band inversion at both the $0$-gap and $\pi$-gap. Specifically, Fig.~\ref{fig_phase_transition} (b) compares numerical solutions with analytical approximations under the perturbative regime of $A/\omega\ll1 $. Notably, the distinct dynamical signatures of band inversion persist even outside this regime, as demonstrated in  Fig.~\ref{fig_phase_transition} (c).

To elucidate this behavior, we analyze the expression in Eq.~\eqref{xt2}, focusing on its energy-gap-dependent denominator and the exponential phase factor. The CoM dynamics of the coupled bands undergoing band inversion exhibit harmonic oscillations governed by the energy gap $\Delta \epsilon = \epsilon_{mq}-\epsilon_{np}$. Crucially, the inversion process induces a sign reversal of the energy gap $\Delta\epsilon\rightarrow -\Delta \epsilon$, which simultaneously modifies both the temporal evolution of the phase factor and the sign of the amplitude in the denominator. This dual reversal directly translates into a phase shift of the oscillatory CoM trajectory, thereby enabling a clear distinction between the dynamical signatures of the pre- and post-inversion phases.

\section{Floquet topological phase Characterization}
\label{sec5}
Since the band inversion associated with the phase transition, as well as the change in the band invariant, can be characterized by the CoM dynamics, the topological invariant of the system can be measured starting from a reference system~\cite{Zhang2020, Uenal2019}. In the high-frequency limit, the system's topological properties reduce to those of its static counterpart, where $(\nu_0, \nu_\pi) = (1, 0)$. The winding numbers of the quasienergy bands—derived from the Floquet operator $\hat U(T,0)$—are then given by~\cite{Fruchart2016} \begin{equation} \label{band_invariant} \nu_{0\pi} = \nu_\pi - \nu_0 = -1, \quad \nu_{\pi 0} = \nu_0 - \nu_\pi = +1, \end{equation} with $\nu_{0\pi}$ ($\nu_{\pi 0}$) representing the winding number of the upper (lower) Floquet band. Note that in one-dimensional systems with chiral symmetry, the Zak phase modulo $2\pi$ is equivalent to the winding number; hence, they can be used interchangeably as bulk topological invariants.

For the extended Hamiltonian, as in the static case, band inversion induces a change in the winding numbers of the bands. This change originates from local massive Dirac points, each contributing $\pm 1/2$ topological charge to the total winding number~\cite{Bernevig2013}. In the phase transition depicted in Fig.~\ref{fig_phase_diagram}(b), the band inversion occurs at the $\pi$ gap, and the relevant band pair near $k_x = 0$ is described locally by a one-dimensional massive Dirac Hamiltonian, as indicated in Fig.~\ref{fig_phase_transition}(a). Given that the lower band has winding number $\nu_{0\pi} = -1$ [Eq.~\eqref{band_invariant}] and there are only two possible Dirac points in the model, we infer that the Dirac point at $k_x = 0$ contributes $-1/2$ to $\nu_{0\pi}$. Consequently, after the band inversion, the invariant of the lower band changes by $+1$, resulting in a final value of $0$. Within the framework of the extended Hamiltonian, the presence of mid-gap states is determined by the topology of the bands below the gap. Thus, we conclude that the invariant of the $0$ gap remains unchanged, while the $\pi$ gap acquires a new invariant $\nu_\pi = 1$.

A similar analysis applies to the phase transition shown in Fig.~\ref{fig_phase_diagram}(a), where the band inversion occurs at the $0$ gap. In this case, two Dirac points are located at $k_x = \pm \arccos(J_1/J_2)$ for large driving amplitude $A$. Owing to inversion symmetry, these Dirac points are related by a parity transformation, and their contributions to the winding number are identical. Prior to the transition, the winding number of the lower band is $\nu_{\pi 0} = +1$; after both Dirac points undergo band inversion, it changes to $-1$. As a result, the $\pi$ gap invariant $\nu_\pi$ remains unaltered, while the $0$ gap invariant becomes $\nu_0 = -1$.

For the above arguments, we have implicitly equated the winding number of the time-evolution operator with the one calculated by the extended Hamiltonian. We now formally establish the equivalence between the winding number of the time-evolution operator and that computed via the extended Floquet Hamiltonian, leveraging the Hamiltonian's inversion symmetry. Under the parity operation, the system satisfies $\sigma_x h(k_x,t) \sigma_x = h(-k_x,t)$, which induces a relationship between eigenstates at $k_x$ and $-k_x$. Specifically, up to a phase factor, this symmetry can be expressed as
 \begin{equation}\label{parity_relation} e^{i\phi_n(k_x)} |n(-k_x)\rangle = \sigma_x |n(k_x)\rangle, \end{equation}
 where $\phi_n(k_x)$ accounts for the phase ambiguity inherent to eigenstate definitions.
The winding number of Floquet bands  is defined as
 \begin{equation} 
 \nu^F_n = \frac{1}{\pi} \int_{-\pi}^{\pi} dk_x \langle n(k_x) | i \partial_{k_x} |n(k_x)\rangle. \label{winding_def} 
 \end{equation}
Substituting the eigenstate decomposition
 \begin{equation} 
 |n(k_x)\rangle = \sum_p |u_n^p(k_x)\rangle \label{eigenstate_decomp}, 
 \end{equation}
 into Eq.~\eqref{winding_def} yields
 \begin{equation}
  \begin{split} 
  \nu^F_n &= \frac{1}{\pi} \sum_{p} \int_{-\pi}^{\pi} dk_x \langle u_n^p(k_x) | i \partial_{k_x} |u_n^p(k_x)\rangle \\
   &\quad + \frac{1}{\pi} \sum_{\substack{p,q \ q\neq0}} \int_{-\pi}^{\pi} dk_x  \langle u_n^p(k_x) | i \partial_{k_x} |u_n^{p+q}(k_x)\rangle, 
  \end{split}\label{winding_split} 
  \end{equation}
 where the first term directly corresponds to the winding number $\nu^{\text{EH}}$ computed from the extended Floquet Hamiltonian. To analyze the second term, we apply the parity relation [Eq.~\eqref{parity_relation}] and the eigenstate orthogonality condition for $\langle n0|nq \rangle = \sum_p\langle u_n^p| u_n^{p+q}\rangle=0 $ for $q\neq0$ (arising from band separation in Floquet theory). Under parity transformation,
 \begin{equation}
 \begin{split}
  &\frac{1}{\pi} \sum_{\substack{p, q\neq0}} \int_{-\pi}^{\pi} dk_x  \langle u_n^p(k_x) | i \partial_{k_x} |u_n^{p+q}(k_x)\rangle \\
   &= -\frac{1}{\pi} \sum_{\substack{p, q\neq0}} \int_{-\pi}^{\pi} dk_x  \langle u_n^p(-k_x) | i \partial_{k_x} |u_n^{p+q}(-k_x)\rangle \\
    &= -\frac{1}{\pi} \sum_{\substack{p,q\neq0}} \int_{-\pi}^{\pi} dk_x  \langle u_n^p(k_x) | i \partial_{k_x} |u_n^{p+q}(k_x)\rangle,
 \end{split}\label{second_term_cancellation} 
 \end{equation}
 where the equality follows from changing variables $k_x\rightarrow -k_x$ and invoking Eq.~\eqref{parity_relation}. This symmetry operation forces the second term in Eq.~\eqref{winding_split} to cancel its own contribution, thereby demonstrating that
 \begin{equation}
 \nu^F_n = \nu_n^{\text{EH}}, 
 \end{equation}
 establishing the equivalence between the Floquet winding number and that of the extended Hamiltonian.

\section{summary}
\label{summary}
In summary,  we have established a fundamental connection between the center-of-mass dynamics of a wave packet and the topological properties of a periodically driven system. By constructing a Floquet perturbation theory within the extended Hilbert space, we derived an analytical expression for the CoM evolution, revealing it as a superposition of Zitterbewegung oscillations at frequencies determined by the Floquet quasienergy gaps. Applying this framework to the driven SSH model, we demonstrated that the CoM dynamics serve as a sensitive indicator of Floquet topological phase transitions. The key insight is that a band inversion during a topological transition closes and reopens a quasienergy gap, which in turn dramatically alters the CoM dynamics. This is manifested as either the dominance of a new low-frequency oscillatory mode or a distinct phase shift in the oscillation, as shown in Figs.~\ref{fig_xt_diagram} and~\ref{fig_phase_transition}. These dynamical signatures persist even beyond the perturbative regime, underscoring their robustness. Furthermore, we have shown how the measured CoM response, when compared to a known reference state~\cite{Zhang2020, Uenal2019}, can be used to deduce changes in the system's bulk topological invariants, as encapsulated in the winding numbers of the Floquet bands~\cite{Fruchart2016,Roy2017,Yao2017}. Our findings bridge the gap between Floquet engineering and quantum dynamics, proposing that CoM measurements—readily accessible in cold-atom~\cite{,Bloch2008,Hasan2022} and photonic~\cite{Ozawa2019} experiments—provide a direct pathway to characterize non-equilibrium topological phases.

\begin{acknowledgments}
This work is supported by NSFC (Grant NO. 12104430) and NSFC (Grant NO. 12504588). 
\end{acknowledgments}

\bibliography{ref_bib}% Produces the bibliography via BibTeX.

\end{document}